\documentclass[a4paper]{jpconf}
\usepackage{graphicx}
\usepackage{epsf}
\usepackage{amsfonts}
\usepackage{amsmath}
\usepackage{bm}
\usepackage{color}
\usepackage{slashed}
\newcommand{\be}{\begin{equation} \nonumber}
\newcommand{\ee}{\end{equation}}
\newcommand{\bea}{\begin{eqnarray} }
\newcommand{\eea}{\end{eqnarray}}
\newcommand{\ba}{\begin{array}}
\newcommand{\ea}{\end{array}}
\newcommand{\bit}{\begin{itemize}}
\newcommand{\eit}{\end{itemize}}
\newcommand{\ben}{\begin{enumerate}}
\newcommand{\een}{\end{enumerate}}

\def\lab{\label}
\def\lan{\langle}

\def\lf{\left}

\def\pa{\partial}
\def\ran{\rangle}
\def\rar{\rightarrow}

\def\ri{\right}

\def\al{\alpha}

\def\de{\delta}

\def\ep{\epsilon}

\def\la{\lambda}
\def\La{\Lambda}
\def\si{\sigma}

\def\om{\omega}

\def\bk{{\bf k}}

\begin{document}

\title{On the canonical quantization of the electromagnetic field and the emergence of gauge invariance }


%
\author{M.~Blasone$^1$, E.~Celeghini$^2$, P.~Jizba$^3$, F.~Scardigli$^{4,5}$ and G.~Vitiello$^1$}
\address{${}^1$ Dipartimento di Fisica, Universit\`a
di Salerno, Fisciano (SA) - 84084 Italy \& INFN, Gruppo Collegato di Salerno}
\address{${}^2$Dipartimento di Fisica, Universit\`a di Firenze and INFN-Firenze,
50019 Sesto Fiorentino, Italy}
\address{${}^3$FNSPE, Czech Technical
University { in Prague}, B\v{r}ehov\'{a} 7, 115 19 Praha 1, Czech Republic} %
\address{${}^4$Dipartimento di Matematica, Politecnico di Milano, Piazza L.da Vinci 32, 20133 Milano, Italy 
} %
\address{${}^5$Institute-Lorentz for Theoretical Physics, Leiden University, P.O. Box 9506, Leiden, The Netherlands}

\begin{abstract}
In the framework of the canonical quantization of the electromagnetic field, we impose as primary condition on the dynamics the positive definiteness of the energy spectrum. This implies that (Glauber) coherent states have to be considered for the longitudinal and the scalar photon fields. As a result  we obtain that the relation holds which in the traditional approach is called the Gupta-Bleuler condition. Gauge invariance emerges as a property of the physical states. The group structure of the theory is recognized to be the one of $SU(2) \otimes SU(1,1)$.
\end{abstract}


%

\section{Introduction}

The quest for a deterministic basis of present physical theories has led 't Hooft to propose a cellular automaton interpretation of Quantum Mechanics \cite{tHbook}: here the central concept is that of a complete set of operators commuting at all times, the {\em beables}, representing the true (ontological) variables of the system and allowing for a deterministic description.
One possibility in such a framework, is represented by loss of information: it may happen that a deterministic theory has a dissipative character, so that, on a given time scale, the dynamics leads to a reduction of the number of degrees of freedom of the system, and the result is the emergence of ordinary quantum states with non-local character.
Several examples of such a scheme have been discussed \cite{tHbook}--\cite{scard2}, in particular the system of two coupled damped--amplified harmonic oscillators (Bateman system)\cite{bateman}--\cite{Schuch}, turned out to be a paradigmatic one, in the sense that the quantum harmonic oscillator here emerges out of the dissipative dynamics of the original (deterministic) system \cite{blasone1,scard}.
A common feature of these models, is the fact that the Hamiltonian of the
(primordial) deterministic system is not positive--definite implying a spectrum unbounded from below: by imposing the constraint of positive definiteness, one then obtains a genuine quantum system with unitary evolution and a stable groundstate.

Motivated by such ideas, we consider in this paper the quantization procedure for the electromagnetic field in Lorenz gauge in the operatorial version, which historically was obtained in the 1950's by Gupta and Bleuler.
We show that, by only resorting to the condition of positive definiteness of the Hamiltonian, i.e. without assuming the gauge invariance as a primary ingredient of the theory,   we arrive at a definition of the physical states of the theory in terms of coherent states of longitudinal and scalar photons, which turn out to be equivalent to the traditional Gupta--Bleuler states.
We also show that the theory has the group structure  of $SU(2) \otimes SU(1,1)$.

The structure of the paper is as follows. In  Section 2, we review the traditional quantization procedure for the electromagnetic field, in Section 3 we present our results. Section 4 is devoted to conclusions and perspectives.

\section{Canonical quantization of electromagnetic field}

In  this Section we summarize briefly the main steps in the canonical quantization of the electromagnetic field in the traditional approach. In our presentation we closely follow  Ref.\cite{greiner}.

Canonical quantization of the Maxwell field in the Lorenz gauge requires the introduction of a gauge fixing term leading to the so-called Fermi Lagrangian density
\cite{greiner}
\bea\label{FermiLag}
\cal{L}&=&-\frac{1}{4}F^{\mu\nu}F_{\mu\nu}\,-\,\frac{1}{2}\zeta \lf(\pa_\mu A^\mu\ri)^2
\eea

Equations of motions following from such Lagrangian are
\bea
\Box A^\mu - (1-\zeta) \pa_\mu \lf(\pa_\si A^\si\ri) &=&0
\eea

If we restrict to the case $\zeta=1$ (Fermi-Feynman gauge), Lagrangian and equations of motion assume the simple form:
\bea
\cal{L}&=&-\frac{1}{2}\pa_\mu A_\nu\pa^\mu A^\nu
\\
\Box A^\mu &=&0
\eea

The Fourier expansion of the $A^\mu$ field can be written as %
\bea \label{field}
A^\mu(x)&=& \int \frac{d^3 k}{\sqrt{2\om_k(2\pi)^3}}\sum_{\la=0}^3\lf( a_{\bk,\la}\,\ep^\mu(\bk,\la)e^{-i k\cdot x} +a^*_{\bk,\la}\,\ep^\mu(\bk,\la)e^{i
k\cdot x} \ri)
\eea

By introducing the conjugate momenta
\bea
\pi_\mu &=&\frac{\pa {\cal L}}{\pa (\pa_0 A^\mu)}
\eea
we obtain the Hamiltonian density
\bea\label{FermiH1}
\cal{H}&=&-\frac{1}{2}\pi_\mu\pi^\mu + \frac{1}{2}\pa_k A_\nu\pa^k A^\nu
\eea
which can be rewritten as
\bea
\cal{H}&=&\frac{1}{2}\sum_{k=1}^2\lf[ (\dot{A^k})^2 + (\nabla A^k)^2\ri]
\, -\,\frac{1}{2}\lf\{\lf[ (\dot{A^0})^2 + (\nabla A^0)^2\ri]-\lf[ (\dot{A^3})^2 + (\nabla A^3)^2\ri]\ri\}.
\eea

Quantization is achieved by imposing commutation relations for the  operators $A^\mu$ and $\pi^\mu$:
\bea
&&\lf[A^\mu({\bf x},t),\pi^\nu({\bf y},t)\ri]\,=\, i g^{\mu \nu} \de^3({\bf x}-{\bf y})
\eea
or
\bea
&&\lf[A^\mu({\bf x},t),\dot{A}^\nu({\bf y},t)\ri]\,=\, i g^{\mu \nu} \de^3({\bf x}-{\bf y})
\eea

By use of  the field expansion (\ref{field}), we obtain the commutation relations for the ladder operators:
\bea\label{acomm}
&&\lf[a_{{\bf k}',\la'},a_{{\bf k},\la}^\dag\ri]\,=\, - g_{\la \la'} \de^3({\bf k}'-{\bf k})
\eea
and the Hamiltonian becomes:
\bea\label{FermiQH1}
H &=&\int d^3{\bf k}\,\om_k\,\lf(\sum_{\la=1,3}a_{{\bf k},\la}^\dag a_{{\bf k},\la} - a_{{\bf k},0}^\dag a_{{\bf k},0} \ri)
\eea

At this point, the Lorenz gauge  is implemented as a condition (the Gupta--Bleuler condition) on the Hilbert space, defining the physical states
$|\Phi\ran $ as those for which %
\bea\label{GBcond1}
\pa^\mu A_\mu^{(+)}|\Phi\ran \, =\, 0
\eea
 where the $``+"$ denotes the positive frequency part of the field. The condition (\ref{GBcond1}) gives
\bea\label{GBcond2}
(a_{{\bf k},0} -a_{{\bf k},3})|\Phi\ran \, =\, 0
\eea
implying that
\bea
\lan \Phi |\lf(a_{{\bf k},0}^\dag a_{{\bf k},0} - a_{{\bf k},3}^\dag a_{{\bf k},3}\ri)|\Phi\ran \,= \, 0.
\eea
%

It can be also shown \cite{greiner} that the physical states can be generated from the purely transverse states $|\Phi_T\ran $ in the following way: %
\bea
|\Phi\ran\, =\, R |\Phi_T\ran
\eea
where
\bea
R \, = \, 1 + \int d^3 k c({\bf k})  L_{\bf k}^\dag +
\int d^3 k d^3 k' c({\bf k}) c({\bf k'})L_{\bf k}^\dag L_{\bf k'}^\dag + \ldots
\eea
and $L_{\bf k}\equiv a_{{\bf k},0} -a_{{\bf k},3}$ with $[L_k,R]=0$. The purely transversal states $|\Phi_T\ran $ are those which do not contain any longitudinal or scalar photons:
\bea
a_{{\bf k},0}|\Phi_T\ran \, = \,a_{{\bf k},3}|\Phi_T\ran \, =\, 0.
\eea
Thus, for such states,  the condition (\ref{GBcond2}) is trivially satisfied.
The physical states $|\Phi\ran $ defined by the Gupta-Bleuler condition $L_{\bf k}|\Phi\ran =0$, are related to the transverse states $|\Phi_T\ran$ through a gauge transformation:
\bea
\lan \Phi |A_\mu(x)|\Phi \ran = \lan \Phi_T |A_\mu(x) + \pa_\mu \La(x)|\Phi_T \ran
= \lan \Phi_T |A_\mu(x) |\Phi_T \ran + \pa_\mu \La(x).
\eea
where  $\La(x)$ is a c-number function satisfying $\Box \La(x) =0$ \cite{greiner}.


\section{Algebraic structure of the  Hamiltonian and coherent states}

We take as starting point the Lagrangian density Eq.(\ref{FermiLag}) with  $\zeta=1$ (Fermi-Feynman gauge):
\bea\label{FermiLag2}
\cal{L}&=&-\frac{1}{4}F^{\mu\nu}F_{\mu\nu}\,-\,\frac{1}{2}  \lf(\pa_\mu A^\mu\ri)^2.
\eea

We stress that we assume now the theory described by Lagrangian
(\ref{FermiLag2}) as fundamental, contrarily to the usual pattern described in previous Section, in which one starts from the Maxwell theory exhibiting gauge invariance
and then introduces the Lagrangian (\ref{FermiLag}) in order to implement canonical quantization. In such a way, we privilege the possibility of a canonical formalism, with respect to  the gauge invariance, which we
do not assume as a fundamental concept.

The Hamiltonian density following from (\ref{FermiLag2}) is the one given in Eq.(\ref{FermiH1}), leading, upon quantization, to the Hamiltonian operator
(\ref{FermiQH1}).

Let us focus  on the algebraic structure of the Hamiltonian (\ref{FermiQH1}). To this end, we define the following operators:
\bea
&& J_+ \,\equiv \, a_{{\bf k},1}^\dag a_{{\bf k},2} \, ,\quad J_- \,\equiv \, a^\dag_{{\bf k},2} a_{{\bf k},1} \, , \quad J_3 \,\equiv\,
\frac{1}{2}\lf(a_{{\bf k},1}^\dag a_{{\bf k},1} - a_{{\bf k},2}^\dag a_{{\bf k},2}\ri) \,
\\[2mm]
&& \lf[ J_+, J_-\ri]\, = \, 2 \, J_3\, ,\quad \lf[ J_3, J_+\ri]\, = \, + \, J_+ \, ,\quad
\lf[ J_3, J_-\ri]\, = \, - \, J_-
\eea
and
\bea
&& K_+ \,\equiv \, a_{{\bf k},3}^\dag a_{{\bf k},0} \, ,\quad K_- \,\equiv \, a^\dag_{{\bf k},0} a_{{\bf k},3} \, ,\quad K_3 \,\equiv\,
\frac{1}{2}\lf(a_{{\bf k},0}^\dag a_{{\bf k},0} + a_{{\bf k},3}^\dag a_{{\bf k},3}\ri) \,
\\ [2mm]
&& \lf[ K_+, K_-\ri]\, = \,- 2 \, K_3 \, ,\quad \lf[ K_3, K_+\ri]\, = \, + \, K_+ \, ,\quad
\lf[ K_3, K_-\ri]\, = \, - \, K_-
\eea
We see that the minus sign in the commutation relations for the $a_0$ operator reflects into the different algebraic structure for the $K$ operators,
closing the  $su(1,1)$ algebra, with respect to the $J$ operators, which close the $su(2)$ algebra.

The Casimir operators are $J_0=\frac{1}{2}\lf(a_{{\bf k},1}^\dag a_{{\bf k},1}+ a_{{\bf k},2}^\dag a_{{\bf k},2}\ri)$ and $K_0=\frac{1}{2}\lf(a_{{\bf
k},0}^\dag a_{{\bf k},0} - a_{{\bf k},3}^\dag a_{{\bf k},3}\ri)$.

In the above definitions, for notational simplicity, we have omitted the $k$ index on the generators $J$ and $K$.

Thus the hamiltonian (\ref{FermiQH1}) can be written as 
\bea \label{HJK}
H &=&\int d^3{ k}\,\om_k\,\lf(J_0 \, - \, K_0  \ri) .
\eea

The fact that the Hamiltonian is written in terms of Casimir operators is remarkable \cite{Iachello} and guarantees the conservation of energy under transformations
induced by the algebraic generators  $J$ and $K$ above defined.

The Hamiltonian (\ref{HJK}) is clearly not positive definite, therefore we  define the physical states as those for which the Hamiltonian reduces to  $J_0$, namely:
\bea\label{tHcond2}
\,_{phys}\lan\psi| K_0|\psi\ran_{phys} \, =\, 0 .
\eea
Our main task is  to
characterize these physical states and analyze if they are equivalent to the physical states defined via the GB condition.

The problem  is to find the explicit form of such physical states. We start by assuming them to be of the form:
\bea\label{Phys}
|\psi\ran_{phys} &= & \prod_{\bf k} |n_{{\bf k},1}\ran_1 \otimes |n_{{\bf k},2}\ran_2 \otimes |\alpha_{\bf k}\ran
\eea
where we denote with subscript 1 and 2 the states for the transverse photons and with $|\alpha_{\bf k}\ran $ a generic
state (to be determined) for the longitudinal and scalar photons.  Although for notational simplicity we do not make it explicit, it is understood that the tensorial product notation used in Eq.~(\ref{Phys}) is also adopted for the operators, e.g. $a_{{\bf k},1}^\dag \equiv a_{{\bf k},1}^\dag \otimes 1 \otimes 1 $, etc..

We now concentrate on the state $|\alpha\ran \equiv\prod_{\bf k} |\alpha_{\bf k}\ran $, which is required to satisfy the above condition (\ref{tHcond2}), i.e.
\bea\lab{tHcond3}
\lan \al |\lf(a_{{\bf k},0}^\dag a_{{\bf k},0} - a_{{\bf k},3}^\dag a_{{\bf k},3}\ri)|\al\ran \, = \, 0.
\eea
We also require such a state to be different from the vacuum state $|0\ran \equiv |0\ran_3\otimes|0\ran_0$, which trivially fulfills such condition.
Furthermore, we restrict to states of the form $|\al\ran = |\al\ran_3\otimes |\al\ran_0$ where $|\al\ran_{3}$ and $|\al\ran_{0}$ denote (Glauber) coherent states for
$a_3$ and $a_0$:
\bea \label{al1}
a_{{\bf k},3} |\al\ran_{3} \, = \, \al_k |\al\ran_{3}
\\ [2mm]
a_{{\bf k},0} |\al\ran_{0} \, = \, \al_k |\al\ran_{0} \label{al2}
\eea
with the same $\al_k$, for any ${\bf k}$.
The coherent state generators are
\bea
&&G_3(\al)=\exp\sum_{\bf k}\lf(\al_k^* \,a_{{\bf k},3} - \al_k\, a_{{\bf k},3}^\dag \ri)
\\
&& |\al\ran_{3}\, = \,G_3^{-1}(\al) |0\ran 
\\ [2mm] \label{coh3}
&& a_{{\bf k},3}(\al) \equiv  G_3^{-1}(\al) a_{{\bf k},3}G_3(\al)\,=\, a_{{\bf k},3} - \al_k\,
\eea
and
\bea
&&G_0(\al)=\exp\sum_{\bf k}\lf(-\al_k^* \,a_{{\bf k},0} + \al_k\, a_{{\bf k},0}^\dag \ri)
\\
&& |\al\ran_{0}\, = \,G_0^{-1}(\al) |0\ran 
\\ [2mm]  \label{coh0}
&& a_{{\bf k},0}(\al)  G_0^{-1}(\al) a_{{\bf k},0}G_\al\,=\, a_{{\bf k},0} - \al_k \, .
\eea
The sign difference in the  commutator for  $a_0$ and $a_0^\dag$ has dictated the choice of the sign in the definition of the $G_0$ generator.
We thus obtain
\bea\lab{GB0a}
\lf( a_{{\bf k},0}-a_{{\bf k},3}\ri)|\al\ran \, = \, 0
\\ [2mm]
\lab{GB0b}
\lan \al|\lf( a^\dag_{{\bf k},0}-a^\dag_{{\bf k},3}\ri) \, = \, 0 \, ,
\eea
which immediately extends to the physical states $|\psi\ran_{phys}$. The condition (\ref{tHcond2}) easily follows from these relations. Note that
Eqs.(\ref{GB0a}), (\ref{GB0b}) are actually the Gupta-Bleuler condition Eq.(\ref{GBcond2}) and its hermitan conjugate.

By resorting to the  $L_{\bf k}$ and $L_{\bf k}^\dag$ operators above defined, we note that they are invariant under  $\al$-translation:
$G_0^{-1}(\al)G_3^{-1}(\al)
\lf( a_{{\bf k},0}-a_{{\bf k},3}\ri)G_3(\al)G(\al)=
\lf( a_{{\bf k},0}-a_{{\bf k},3}\ri)$ and similar for
$L_{\bf k}^\dag$.

Let us now consider the explicit form of the coherent states $|\al\ran_0$ and $|\al\ran_3$. By use of the Campbell-Baker-Hausdorff relation and the
commutation relations Eq.(\ref{acomm}), we obtain
\bea
|\al\ran_3&=&
\exp\lf(-\frac{1}{2} \int d^3k \,|\al_k|^2\ri)
\exp\lf(\int d^3k \,\al_k \,a_{{\bf k},3}^\dag\ri)
|0\ran_3
\\
|\al\ran_0&=&
\exp\lf(\frac{1}{2} \int d^3k\, |\al_k|^2\ri)
\exp\lf(-\int d^3k\, \al_k \,a_{{\bf k},0}^\dag\ri)
|0\ran_0
\eea
and
\bea
|\al\ran &\equiv &|\al\ran_3 \otimes |\al\ran_0=
\exp\lf(\int d^3k\, \al_k \,\lf(a_{{\bf k},3}^\dag -a_{{\bf k},0}^\dag\ri) \ri)
|0\ran
\\
&=&\lf(1 + \int d^3k \,(-\al_k) \,L_{\bf k}^\dag +
\int d^3k d^3k' \,\frac{(-\al_k)(-\al_{k'})}{2!}
L_{\bf k}^\dag L_{\bf k'}^\dag \, + \, ...
\ri)|0\ran
\eea

Thus a one-to-one correspondence exists between the coherent states above defined and those used in the Gupta-Bleuler quantization. We can therefore
identify the physical states of the Gupta-Bleuler condition with the ones defined by the condition Eq.(\ref{tHcond2}).

We can thus write
\bea
\,_{phys}\lan\psi|A_\mu (x)|\psi\ran_{phys}
 &=&
\lan \phi_T|A_\mu (x)|\phi_T\ran \,+ \,\pa_\mu \La(x)
\eea
with $\La(x)$ satisfying $\Box \La =0$.

In conclusion, only the transverse photons contribute to the energy in the coherent state representation $\{ |\psi\ran_{phys}\}$ :
\bea
\,_{phys}\lan\psi|H |\psi\ran_{phys}&=&\int d^3{ k}\,\om_k\,\sum_{\la=1,2}n_{{\bf k},\la}.
\eea

We conclude that, by choosing the coherent state representation $|\al\ran $, we obtain Eq.(\ref{tHcond3}) which in the Gupta-Bleuler method is instead
postulated in order to impose the Lorenz gauge condition. In our case, we do not assume any gauge invariance from the beginning, rather we obtain it as an emergent property of the physical states.

\section{Conclusions}

In this paper we have explored the possibility to consider as fundamental the theory described by the Lagrangian density Eq.(\ref{FermiLag2}). On the other hand, the traditional approach is to start with the gauge invariant Maxwell theory and then proceed to the canonical quantization, the gauge invariance being assumed to be a fundamental ingredient of the theory. In our approach the canonical structure, namely the operatorial algebraic structure of the theory, is instead assumed to be fundamental, the gauge invariance being implied by the positive definiteness of the Hamiltonian operator. In short, we show that the primary condition that the energy spectrum has to be definite positive fully determines the structure of the physical states of the theory. Thus, instead of postulating it, we obtain as a result what in the traditional approach is called the Gupta-Bleuler condition. 

Remarkably, the Hamiltonian for the electromagnetic field is shown to be given by the Casimir operators of the $su(2)\otimes su(1,1)$ algebras defining the theory. Such a result is a general feature of the dynamics, holding in the traditional approach and in our treatment, and makes evident the invariance of the spectrum under the  of $SU(2) \otimes SU(1,1)$ group transformations.

Finally, note that by acting with the generator $G_3$ and $G_0$ on the $K_0$ term in the Hamiltonian in Eq.(\ref{HJK}), instead than on the states  $|0\ran_{3}$ and $|0\ran_{0}$, we have
\bea
a_{{\bf k},3}^\dag a_{{\bf k},3} - a_{{\bf k},0}^\dag a_{{\bf k},0} \rar \lf( a_{{\bf k},3}^\dag a_{{\bf k},3} - {\al^*_k} a_{{\bf k},3} - \al_k  a_{{\bf k},3}^\dag \ri) \, - \, \lf(  a_{{\bf k},0}^\dag a_{{\bf k},0} - \al^*_k a_{{\bf k},0} \, - \, \al_k  a_{{\bf k},0}^\dag \ri) \, ,
\eea
i.e., under the shift transformations Eqs.~(\ref{coh3}) and (\ref{coh0}), the longitudinal and the scalar photon modes are submitted to external forces (represented by the linear operator terms $a_{{\bf k},3}$  and $a_{{\bf k},0}$ and their hermitian conjugates, respectively). This shows that the two sets of longitudinal and scalar modes, each constitutes actually an open dissipative system on which an external force acts. The energy of each of the modes, for given $\bf k$, is proportional to  $|\al_k |^2$ (cf.  Eq.~(\ref{tHcond3})). Both sets, combined together, form a closed system whose total energy is conserved and vanishes.

It is in our future plans to extend the analysis here presented in the direction of the 't Hooft scheme for quantization of deterministic systems \cite{tHbook}. In this context, the emergence of various symmetries, including local gauge symmetry, has been discussed in Refs.\cite{tHooft:2007nis,Elze}. Another development will be the formulation of these topics in the path-integral formalism, in the line of Ref.\cite{BJK1}.

\section*{Acknowledgements}
P.J.  was  supported  by the Czech  Science  Foundation Grant No. 17-33812L. M.B., E.C. and G.V. were supported by INFN and Miur.

\section*{References}


\begin{thebibliography}{99}

%
\bibitem{tHbook}
't Hooft G 2016 {\it The Cellular Automaton Interpretation of Quantum Mechanics} (Springer Int. Publ.)

%
%

\bibitem{blasone1} Blasone M, Jizba P and Vitiello G 2001 {\it Phys. Lett. A} {\bf 287} 205


%
%
%
\bibitem{blasone2}
Blasone M, Celeghini E, Jizba P and Vitiello G  2003
  {\it Phys. Lett. A} {\bf 310}  393.
%


%
\bibitem{scard}
Blasone M, Scardigli F, Jizba P and Vitiello G
2009 {\it Phys. Lett. A} {\bf 373} 4106.
%

\bibitem{scard2}
Scardigli F 2007 {\it Found. of Phys.}, {\bf 37}, 1278.
%

\bibitem{bateman} Bateman H 1931 {\it Phys. Rev.} {\bf 38} 815

%
\bibitem{celeghini92} Celeghini E, Rasetti M and Vitiello G 1992 {\it Ann. Phys.} {\bf 215}  156.



\bibitem{blasone3} Blasone M and Jizba P 2004  {\it Ann. Phys.} {\bf 312} 354.
Blasone M, Jizba P, Vitiello G 2003 {\it J. Phys. Soc. Jap.} {\bf  72 }(suppl. C)  50;
Blasone M, Jizba P 2002 {\it Can. J. Phys.} {\bf 80} 645.

%
\bibitem{chruscinski}Chruscinski D 2006 {\it Ann. Phys.} {\bf 321} 840; Chruscinski D and Jurkowski J 2006 {\it Ann. Phys.} {\bf 321} 854.
%

\bibitem{Schuch}
 Schuch D  and Blasone M 2017
  {\it J.\ Phys.\ Conf.\ Ser.}  {\bf 880} 012050.
 
\bibitem{greiner}
Greiner W and Reinhardt J 1996
{\em Field Quantization} (Springer-Verlag, Berlin)




\bibitem{Iachello}
Iachello F 2015  {\em Lie Algebras and Applications} (Lecture Notes in Physics, Springer)




\bibitem{tHooft:2007nis}
 't Hooft G.~2007
  {\it AIP Conf.\ Proc. } {\bf 957}   154

\bibitem{Elze}
Elze H T 2009
  {\it J.\ Phys.\ Conf.\ Ser. } {\bf 171}   012034.

%
\bibitem{BJK1}
Blasone M, Jizba  P and Kleinert H 2005 {\it Phys. Rev.} A {\bf 71} 052507;
2005 {\it Ann. Phys.} {\bf 320}  468.
%

\end{thebibliography}
\end{document}